\documentclass{appolb-blank}  

\usepackage{graphicx}

\usepackage{latexsym,amssymb,amsmath,amsfonts,epsfig,mathtools}  

\newcommand{\beq}{\begin{equation}}   
\newcommand{\eeq}{\end{equation}}
\newcommand{\beqa}{\begin{eqnarray}}
\newcommand{\eeqa}{\end{eqnarray}}
\newcommand{\beqNO}{\begin{equation*}}
\newcommand{\eeqNO}{\end{equation*}}
\newcommand{\beqaNO}{\begin{eqnarray*}}
\newcommand{\eeqaNO}{\end{eqnarray*}}
\newcommand{\bsubeqs}{\begin{subequations}}
\newcommand{\esubeqs}{\end{subequations}}

\begin{document}

\eqsec  

\title{Quantum scattering process and information transfer\\
 out of a black hole
 \\}   

\author{V.A. Emelyanov, F.R. Klinkhamer
\address{
Institute for Theoretical Physics,
Karlsruhe Institute of Technology (KIT),\\
76128 Karlsruhe, Germany\\
\texttt{viacheslav.emelyanov@kit.edu, frans.klinkhamer@kit.edu}
}\\ 
}
\maketitle
\begin{abstract}
We calculate the probability amplitude
for tree-level elastic electron-muon scattering
in Minkowski spacetime
with carefully prepared initial and final wave packets.  
The obtained nonzero amplitude implies a nonvanishing
probability for detecting a recoil electron outside
the light-cone of the initial muon.
Transposing this Minkowski-spacetime scattering result
to a near-horizon
spacetime region
of a massive Schwarzschild black hole and referring to a
previously proposed \textit{Gedankenexperiment}, we conclude
that, in principle,
it is possible to have information transfer from inside  
the black-hole horizon to outside.
\end{abstract}

\vspace{-125mm}
\noindent Acta Phys. Pol. B 52, 805 (2021) \hfill arXiv:2010.15787
\vspace{+125mm}



\section{Introduction}
\label{sec:Introduction}

It is often said that nothing can come out of a black hole
(cf. Sec.~33.1 of Ref.~\cite{MisnerThorneWheeler2017}).
This statement may, however, be not quite correct,
as we have argued previously~\cite{EmelyanovKlinkhamer2018}
that \emph{information} can come out.
The information is carried
not by a particle but by momentum
transfer in a quantum scattering process.
In fact, the information-transfer process is
based \emph{not} on a faster-than-light signal exchange but
on a \emph{virtual} photon exchange. This leads to
dynamical entanglement between two initially unentangled
charged particles which are located at different sides of the
black-hole horizon.
A \textit{Gedankenexperiment}~\cite{EmelyanovKlinkhamer2018}
relying on such momentum transfer
(or its absence, i.e., no quantum scattering)
allows, in principle, for the transmittal of
an elementary message (bit value ``1'' or ``0'') from inside
the black-hole horizon to outside the black-hole horizon.

\begin{figure}[t] 
\begin{center}
\includegraphics[scale=1.1]{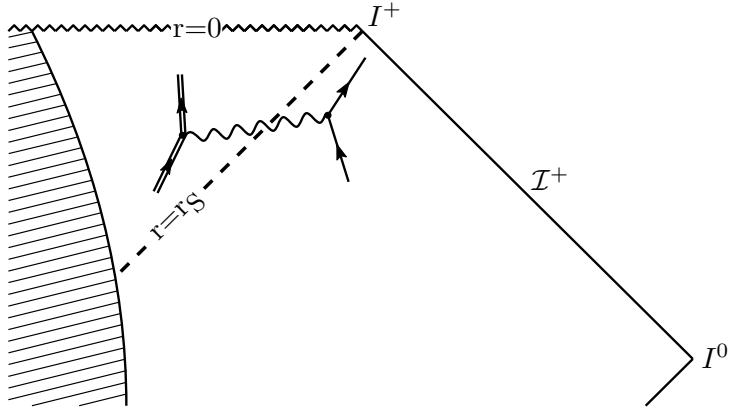}
\end{center}
\vspace*{-0mm}
\caption{Sketch of an elastic electron-muon scattering process
in the near-horizon region of a massive Schwarzschild black hole.
Shown is part of the Penrose conformal diagram of
the Schwarzschild black hole in Kruskal--Szekeres coordinates 
(see Figs.~34.3 and 32.1.b  of Ref.~\cite{MisnerThorneWheeler2017} for details).
The Schwarzschild horizon $r=r_S \equiv 2 G M_\text{BH}/c^2$
is indicated by the dashed line
and the matter of a spherically-symmetric collapsing star by
the shaded region on the left of the figure.
The scattering process is shown symbolically by
a position-space Feynman diagram~\cite{Feynman1949} with
a single line for the electron, a double line for the muon,
and a wavy line for the exchange photon.
}
\label{fig:bh-and-alcs}
\end{figure}

The goal of the present paper is to present a straightforward
quantum-electrodynamics~\cite{Feynman1949,Veltman1994}
calculation
that indisputably shows the nonvanishing probability for having
such a momentum transfer.
Specifically, we consider a carefully prepared $2$--$2$ scattering
process in Minkowski spacetime.
According to the Einstein Equivalence Principle,
this Minkowski-spacetime scattering process
is relevant for the near-horizon spacetime region
of a large-enough black hole;
see Fig.~\ref{fig:bh-and-alcs} for a sketch.
This near-horizon region with local inertial coordinates
is described by a patch of Minkowski spacetime
with a projected black-hole horizon that
corresponds to part of a light-cone
(see Fig.~2 in Ref.~\cite{EmelyanovKlinkhamer2018}).

The setup of our $2$--$2$ scattering process in
Minkowski spacetime involves a muon $\mu^{-}$
and an electron $e^{-}$, where the initial muon
is strictly localized inside the projected black-hole horizon and
the initial electron, localized with large probability
outside the projected black-hole horizon,
has strictly an incoming 3-momentum
(i.e., 3-momentum directed towards the black hole center).
The aim of the calculation is to establish a nonzero probability
for finding, at a later time, an \emph{outgoing} electron
(3-momentum directed away from the black hole center) by use of
a large detector positioned outside the projected black-hole horizon.

Let us immediately put to rest
possible worries about causality.
The setup of the Minkowski-spacetime
scattering process is such that there is
an \emph{extended} initial state and the
final recoil electron lies
\emph{within} the outermost light cone of
the initial electron, so that there is no problem with causality.
As mentioned above, the final electron-muon state is
entangled, whereas the initial electron-muon state
is not~\cite{EmelyanovKlinkhamer2018}.

The outline of our paper is as follows.
In Sec.~\ref{sec:Free-wave-packets-Minkowski-spacetime}, we describe
carefully chosen wave packets in Minkowski spacetime
for the initial and final muons and electrons.
In Sec.~\ref{sec:Quantum-scattering-Minkowski-spacetime},
we calculate the scattering probability amplitude
of these initial and final particles in quantum electrodynamics
and obtain a nonzero probability amplitude for detecting a
recoil electron outside the light-cone of the initial muon.
(The scattering-probability-amplitude calculation
of the present paper improves upon the one
of App.~C in Ref.~\cite{EmelyanovKlinkhamer2018}
precisely by the detailed discussion of appropriate
initial and final wave packets.) 
In Sec.~\ref{sec:Across-horizon-scattering-for-massive-black-hole},
we transpose the Minkowski-spacetime result to
a near-horizon spacetime region of a massive Schwarzschild black hole
and recall the essential steps of a
\textit{Gedankenexperiment} from our
previous paper~\cite{EmelyanovKlinkhamer2018}, which then
allows for information transfer from inside the
black-hole horizon to outside.

It is possible, in a first reading, to skip
the technical details and to move immediately
to Secs.~\ref{subsec:Numerical-results-probability-amplitude}
and \ref{sec:Across-horizon-scattering-for-massive-black-hole}.


\noindent
\section{Free wave packets in Minkowski spacetime}
\label{sec:Free-wave-packets-Minkowski-spacetime}

\subsection{General wave-packet solution}
\label{subsec:General-wave-packet-solution}

We consider a general wave-packet solution of the \emph{free}
Dirac equation for mass $M$,
with center-of-mass parameters $X^a = (X^0,\mathbf{X})$
in position space and $P^a = (P^0,\mathbf{P})$ in momentum space
(the localization regions satisfy the Heisenberg
uncertainty relations).
With gamma matrices in the Weyl representation,
this solution reads
\bsubeqs\label{eq:wave-function}
\beqa
\hspace*{-5mm}
\psi_{X,P}(x) &\equiv& {\int}\frac{d^4Q}{(2\pi)^3}\,
\theta\big(Q^0\big)\,\delta\big(Q^2-M^2\big)F_P(Q)\,u(Q)\,e^{-iQ{\cdot}(x-X)}\,,
\eeqa
where spinor indices have been suppressed and
\beqa
u(Q) &=& \left(
\begin{array}{c}
\sqrt{Q{\cdot}\sigma}\,\xi \\[1mm]
\sqrt{Q{\cdot}\bar{\sigma}}\,\xi \\
\end{array}
\right),
\eeqa
\esubeqs
with a two-component spinor $\xi$ normalized by $\xi^\dagger\xi = 1$.
We shall assume in what follows that $\xi^T = (1,\,0)$.
Furthermore, $\sigma^{a} \equiv  \big(\sigma^0,\,+\sigma^i\big)$
and $\bar{\sigma}^{a} \equiv \big(\sigma^0,\,-\sigma^i\big)$ are
two matrix-valued four-vectors with $\sigma^0$ and $\sigma^i$ standing
for the $2{\times}2$ identity matrix and the three Pauli matrices.

With a momentum wave function $F_P(Q)$ peaking at $Q = P$,
the normalized quantum state $|\psi_{X,P}\rangle$
associated with the position wave function $\psi_{X,P}(x)$
from \eqref{eq:wave-function} has
\beqa\label{eq:normalization}
\langle \psi_{X,P}|\psi_{X,P} \rangle &\equiv&
\int_{\mathbb{R}^3} d^3\mathbf{x}\,
\big(\psi_{X,P}(x)\big)^\dagger\,\psi_{X,P}(x)
\nonumber\\[1mm]
&=& \frac{1}{2}{\int}\frac{d^3\mathbf{Q}}{(2\pi)^3}\,
\frac{|F_{P}(\mathbf{Q})|^2}{\sqrt{|\mathbf{Q}|^2 + M^2}}
\;=\; 1\,.
\eeqa

Throughout this paper, we use natural units with $c=1$ and $\hbar=1$.
For standard Cartesian coordinates,
\bsubeqs\label{eq:Cartesian coordinate-Minkowski-metric}
\beqa
x^{a} &=&
\Big[\big(x^{0},\, \mathbf{x}\big)\Big]^{a}
=
\Big[\big(x^{0},\, x^{1},\, x^{2},\, x^{3}\big)\Big]^{a}
\,,
\eeqa
the Minkowski metric reads
\beqa
\label{eq:Minkowski-metric}
\eta_{ab} &=& \Big[\text{diag}\big(1,\, -1,\, -1,\, -1\big)\Big]_{ab}\,,
\eeqa
\esubeqs
where the spacetime indices $a,\,b$ run over $\{0,\, 1,\, 2,\, 3 \}$.
Occasionally, we will write
\beq
x^{0} = c\,t = t\,.
\eeq

\subsection{Muon wave packets}
\label{subsec:Muon-wave-packets}

\subsubsection{Muon-wave-packet Ansatz}
\label{subsubsec:Muon-wave-packet Ansatz}

We take the following \textit{Ansatz} for the momentum
wave function of the muon ($\mu$):
\beqa\label{eq:f-muon}
\big(F_P(Q)\big)_\mu &=&
N_\mu\,2\,Q^0\,
\prod_{i = 1}^{i = 3}\,\text{sinc}\left(\frac{Q_i-P_i}{D}\right)\,,
\eeqa
where $N_\mu$ is the normalization factor, $D$ the momentum variance,
and `$\text{sinc}$' the standard function
\beqa\label{eq:def-sinc}
\text{sinc}(x) \equiv \frac{\sin\,x}{x}\,,
\eeqa
for $x\in \mathbb{R}$.

In order to guarantee that the undisturbed wave packet essentially
propagates as a free classical particle, we assume that
the particle Compton wavelength is much smaller than the packet
localization~size (see below).
Taking $M_\mu \gg D$ in
\eqref{eq:f-muon},
we obtain for the muon wave packet from \eqref{eq:wave-function}
\beqa\label{eq:def-approx-muon-WF}
\hspace*{-5mm}
\big(\psi_{X,P}(x)\big)_\mu &\approx&
\mu_{X,P}(x) \nonumber\\[1mm]
\hspace*{-5mm}
&\equiv&
\frac{N_\mu}{(2/D)^3}\,u(P)\,e^{-iP{\cdot}\Delta{x}}
\prod_{i = 1}^{i = 3}
\,\text{rect}\left(\frac{P^i\Delta{x}^0 - P^0\Delta{x}^i}{2P^0/D}\right),
\eeqa
where
\beqa\label{eq:def-Delta-x-mu}
\Delta{x}^\mu &\equiv& x^\mu - X^\mu\,,
\eeqa
and
where the top-hat function `$\text{rect}(x)$', for $x\in \mathbb{R}$,
is defined as follows:
\beqa\label{eq:def-rect}
\text{rect}(x) &\equiv&
\left\{
\begin{array}{lcll}
1\,, & |x| &<& 1/2\,, \\[1mm]
1/2\,, & |x| &=& 1/2\,, \\[1mm]
0\,, & |x| &>& 1/2\,.
\end{array}
\right.
\eeqa
The right-hand side of \eqref{eq:def-approx-muon-WF}
has nonvanishing support over a \emph{finite} spatial region,%
\beqa
\Delta{x}^i &\in&
\left[P^i\Delta{x}^0/P^0 - 1/D,\,P^i\Delta{x}^0/P^0 +1/D  \right]\,,
\eeqa
which will be an important input for the setup to be discussed
in the rest of this section.

\subsubsection{Muon-wave-packet normalization and boundary conditions}
\label{subsubsec:Muon-wave-packets-normalization}

The normalization factor $N_\mu$ can now be directly computed
with $\mu_{X,P}(x)$ from~\eqref{eq:def-approx-muon-WF}.
Specifically, we find that
\beq\label{eq:normalisation-analytical}
\hspace*{-0mm}
N_{\mu,\,\text{approx.}} = \sqrt{\frac{4}{P_{0}\,D^3}}
\; \Rightarrow \;
N_{\mu,\,i,\,\text{approx.}} = N_{\mu,\,f,\,\text{approx.}} \approx
\frac{0.036965}{D^2}\,,
\eeq
where we assume that
\beqa
M_\mu &=&  2070\,D
\eeqa
and that the initial ($i$) and final ($f$) conditions
for the muon wave packets
are, respectively,%
\bsubeqs\label{eq:ifc-muon}
\beqa
(X)_{\mu,\,i}\,D &\approx& \big(0.00,\,-3.54,\,0.00,\,0.00\big)\,,
\\[2mm]
(P)_{\mu,\,i}\big/D &\approx& \big(2927.42,\,2070.00,\,0.00,\,0.00\big)\,,
\eeqa
and
\beqa
(X)_{\mu,\,f}\,D &\approx& \big(10.00,\,-2.04,\,2.89,\,0.00\big)\,,
\\[2mm]
(P)_{\mu,\,f}\big/D &\approx& \big(2927.42,\,-1195.12,\,1690.15,\,0.00\big)\,.
\eeqa
\esubeqs
The specific numerical values for $(X)_{\mu,\,i}$
and $(X)_{\mu,\,f}$ have been chosen, so that
the position of the undisturbed initial and final muon wave packets  
exactly coincide at $t = 5/D$ with center-of-mass
position $(0,\,0,\,0)$. The boundary conditions \eqref{eq:ifc-muon} 
are appropriate for the quantum scattering process to be discussed in
Sec.~\ref{sec:Quantum-scattering-Minkowski-spacetime},
with initial and  final times
\beq
\label{eq:ti-tf-num}
t_i =0\,,\quad t_f =10/D\,.
\eeq

By inserting $\big(F_P(Q)\big)_\mu$ from \eqref{eq:f-muon}
into~\eqref{eq:normalization} we obtain numerically
\beqa\label{eq:normalisation-numerical}
N_{\mu,\,i,\,\text{num.}} &\approx& N_{\mu,\,f,\,\text{num.}} \;\approx\;
\frac{0.036981}{D^2}\,.
\eeqa
The agreement between~\eqref{eq:normalisation-analytical}
and~\eqref{eq:normalisation-numerical} is better than $0.05\%$.

\subsubsection{Muon-wave-packet motion}
\label{subsubsec:Muon-wave-packets-motion}

We now wish to show that the wave packet $\mu_{X,P}(x)$
essentially propagates like a free classical particle. We find
\bsubeqs\label{eq:muon-motion}
\beqa
\langle\mu_{X,P}|\,\mathbf{x}\,|\mu_{X,P}\rangle_{t} &\equiv&
{\int_{x^0=t}}d^3\mathbf{x}\,\big(\mu_{X,P}(x)\big)^\dagger\,\mathbf{x}\,\mu_{X,P}(x)
\nonumber \\[1mm] &=&
\mathbf{X} +  \mathbf{V}\,\Big(t-X^0\Big)\,,
\\[2mm]
\mathbf{V} &\equiv& \frac{\mathbf{P}}{P^0} \,.
\eeqa
\esubeqs
The result \eqref{eq:muon-motion} corresponds to the trajectory
of a free classical particle in special relativity.

\begin{table*}[t]
\vspace*{0mm}
\begin{center}
\caption{\label{tab:muon-wave-packets}%
Free muon wave packets: motion in three spatial dimensions.}\vspace*{4mm}
\renewcommand{\tabcolsep}{0.8pc} 
\renewcommand{\arraystretch}{1.8} 
\begin{tabular}{ |c|c|c|c|}
\hline
\hline
\multicolumn{4}{|c|}{Initial muon ($t=t_{i}$)} \\
\hline
&
$M_\mu/D=1$ &
$M_\mu/D=10$ &
$M_\mu/D=2070$ \\
\hline
$\left|\langle x^{1} \rangle {-} X^{1} \right|\,D$ & $0$ & $0$ & $0$ \\
$\left|\langle x^2 \rangle {-} X^2 \right|\,D$ &
$\approx 2{\times}10^{-2}$ & $\approx 10^{-2}$ & $\approx 7{\times}10^{-5}$ \\
$\left|\langle x^3 \rangle {-} X^3 \right|\,D$ & $0$ & $0$ & $0$ \\
\hline
$\left|\langle v^{1} \rangle {-} V^{1} \right|$ &
$\approx 0.6$ & $\approx 0.1$ & $\approx 5{\times}10^{-5}$ \\
$\left|\langle v^2 \rangle {-} V^2 \right|$ & $0$ & $0$ & $0$ \\
$\left|\langle v^3 \rangle {-} V^3 \right|$ & $0$ & $0$ & $0$ \\
\hline
\hline
\end{tabular}
\end{center}
\vspace*{0mm}
\begin{center}
\renewcommand{\tabcolsep}{0.8pc} 
\renewcommand{\arraystretch}{1.8} 
\begin{tabular}{ |c|c|c|c| }
\hline
\hline
\multicolumn{4}{|c|}{Final muon ($t=t_{f}$)} \\
\hline
&
$M_\mu/D=1$ &
$M_\mu/D=10$ &
$M_\mu/D=2070$ \\
\hline
$\left|\langle x^{1} \rangle {-} X^{1} \right|\,D$ & $\approx 10^{-2}$ &
$\approx 9{\times}10^{-3}$ & $\approx 6{\times}10^{-5}$ \\
$\left|\langle x^2 \rangle {-} X^2 \right|\,D$ &
$\approx 9{\times}10^{-3}$ & $\approx 6{\times}10^{-3}$ & $\approx 4{\times}10^{-5}$ \\
$\left|\langle x^3 \rangle {-} X^3 \right|\,D$ &
$0$ & $0$ & $0$ \\
\hline
$\left|\langle v^{1} \rangle {-} V^{1} \right|$ &
$\approx 0.3$ & $\approx 0.1$ & $\approx 3{\times}10^{-5}$ \\
$\left|\langle v^2 \rangle {-} V^2 \right|$ &
$\approx 0.5$ & $\approx 0.1$ & $\approx 4{\times}10^{-5}$ \\
$\left|\langle v^3 \rangle {-} V^3 \right|$ & $0$ & $0$ & $0$ \\
\hline
\hline
\end{tabular}
\end{center}
\end{table*}

If we do not approximate the exact wave packet
$\big(\psi_{X,P}(x)\big)_\mu$ by $\mu_{X,P}(x) $,
then we obtain the numerical results shown in Table~\ref{tab:muon-wave-packets}.
Here, we have used the following expressions:
\bsubeqs\label{eq:expectation-value-x-expectation-value-v}
\beqa
\langle \mathbf{x} \rangle_{t} &\equiv&
{\int_{x^0=t}}d^3\mathbf{x}\;
\big(\psi_{X,P}(x)\big)^\dagger\,\mathbf{x}\,\psi_{X,P}(x)\,,
\\[2mm]
\langle \mathbf{v}\rangle_{t}  &\equiv&
\frac{d}{dt}\,\left({\int_{x^0=t}}d^3\mathbf{x}\;
\big(\psi_{X,P}(x)\big)^\dagger\,\mathbf{x}\,\psi_{X,P}(x)\right)\,,
\eeqa
\esubeqs
where the particle label ``$\mu$'' on the wave functions has
been temporarily removed.
Obviously, the expectation values
$\langle \mathbf{x} \rangle$ and $\langle \mathbf{v} \rangle$
from Table~\ref{tab:muon-wave-packets}, where the
suffixes $t_{i}$ and $t_{f}$ have been omitted,
are close to their classical values
for the chosen initial and final conditions,
provided $M_\mu/D \gg 10$.
In Sec.~\ref{sec:Quantum-scattering-Minkowski-spacetime},
we shall make numerical computations with $M_\mu/D = 2070$.

\subsection{Electron wave packets}
\label{subsec:Electron-wave-packets}

\subsubsection{Electron-wave-packet Ansatz}
\label{subsubsec:Electron-wave-packet Ansatz}

We take the following \textit{Ansatz} for the momentum
wave function of the electron ($e$):
\beqa\label{eq:f-electron}
\big(F_P(Q)\big)_e &=&
N_e\,2\,Q^0\,
\prod_{i = 1}^{i = 3}\,\text{rect}\left(\frac{Q_i-P_i}{2\,D}\right)\,,
\eeqa
where $N_e$ is the normalization factor
and the `$\text{rect}$' function has been  defined by \eqref{eq:def-rect}.
Note that we assume, for simplicity, that
the momentum variance $D$ for the muon and electron wave
functions are equal.

Taking $M_e \gg D$ in \eqref{eq:f-electron},
we obtain for the electron wave packet from \eqref{eq:wave-function}
\beqa\label{eq:def-approx-electron-WF}
\hspace*{-7mm}
\big(\psi_{X,P}(x)\big)_e &\approx&
e_{X,P}(x)
\nonumber\\[1mm] &\equiv&
\frac{N_e}{(\pi/D)^3}\,u(P)\,e^{-iP{\cdot}\Delta{x}}
\prod_{i = 1}^{i = 3}
\,\text{sinc}\left(\frac{P^i\Delta{x}^0 - P^0\Delta{x}^i}{P^0/D}\right),
\eeqa
with $\Delta{x}^\mu $ defined by \eqref{eq:def-Delta-x-mu}.
This wave packet has finite support in momentum space and infinite
support in position space. With appropriate $P^{1}$ and $D$
in \eqref{eq:f-electron}, it is possible to get an initial electron
with only negative momenta $Q^{1}$
and a final electron with only positive momenta $Q^{1}$.

\subsubsection{Electron-wave-packet normalization and boundary conditions}
\label{subsubsec:Electron-wave-packet-normalization}

The normalization factor $N_e$ can be calculated with $e_{X,P}(x)$
from \eqref{eq:def-approx-electron-WF},
\beq\label{eq:normalisation-analytical-electron}
\hspace*{-0mm}
N_{e\,\text{approx.}} = \sqrt{\frac{\pi^3}{2\,P_{0}\,D^3}}
\; \Rightarrow \;
N_{e,\,i,\,\text{approx.}} = N_{e,\,f,\,\text{approx.}} \approx
\frac{0.086541}{D^2}\,,
\eeq
where we assume that
\beqa
M_e &=& 10\,D
\eeqa
and that the initial ($i$) and final ($f$) conditions
for the electron wave packets are, respectively,%
\bsubeqs\label{eq:ifc-electron}
\beqa
(X)_{e,\,i}\,D &\approx& \big(0.00,\,2.89+D \, L,\,-4.08,\,0.00\big)\,,
\\[2mm]
(P)_{e,\,i}/D &\approx& \big(2070.02,\,-1195.12,\,1690.15,\,0.00\big)\,,
\eeqa
and
\beqa
(X)_{e,\,f}\,D &\approx& \big(10.00,\,4.99+D \, L,\,0.00,\,0.00\big)\,,
\\[2mm]
(P)_{e,\,f}/D &\approx& \big(2070.02,\,2070.00,\,0.00,\,0.00\big)\,.
\eeqa
\esubeqs
Note that $(X)_{e,\,i}$ and $(X)_{e,\,f}$ have been chosen, so
that the position of the
undisturbed initial and final electron wave packets  
exactly overlap at $t = 5/D$ with center-of-mass coordinates $(L,\,0,\,0)$.
Again, the boundary conditions \eqref{eq:ifc-electron} are appropriate for the
quantum scattering process to be discussed in
Sec.~\ref{sec:Quantum-scattering-Minkowski-spacetime},
with initial and  final times \eqref{eq:ti-tf-num}.

By inserting $\big(F_P(Q)\big)_e$ from \eqref{eq:f-electron}
into~\eqref{eq:normalization} we obtain numerically
\beqa\label{eq:normalisation-numerical-electron}
N_{e,\,i,\,\text{num.}} &\approx& N_{e,\,f,\,\text{num.}} \;\approx\;
\frac{0.086542}{D^2}\,.
\eeqa
The agreement between~\eqref{eq:normalisation-analytical-electron}
and~\eqref{eq:normalisation-numerical-electron} is roughly $0.001\%$.

\subsubsection{Electron-wave-packet motion}
\label{subsubsec:Electron-wave-packet-motion}

Let us now show that the wave packet $e_{X,P}(x)$ essentially propagates
like a free classical particle. We find
\beqa
\langle e_{X,P}|\,\mathbf{x}\,|e_{X,P}\rangle_{t} &\equiv&
{\int_{x^0=t}}d^3\mathbf{x}\,\big(e_{X,P}(x)\big)^\dagger\,\mathbf{x}\, e_{X,P}(x)
\nonumber \\[1mm] &=&
\mathbf{X} + \frac{\mathbf{P}}{P^0}\,\Big(t-X^0\Big)\,,
\eeqa
which corresponds to the trajectory of a free classical particle
in special relativity.

If we do not approximate the exact wave packet
$\big(\psi_{X,P}(x)\big)_e$ by $e_{X,P}(x) $, then
we obtain the numerical results shown in Table~\ref{tab:electron-wave-packets},
where we have used
expressions \eqref{eq:expectation-value-x-expectation-value-v} applied to 
the electron wave functions.
For the case of the electron, $\langle \mathbf{v} \rangle$ is close to
$\mathbf{V}$ for all values of $M_e/D$ considered,
because the electron is relativistic. This implies that the wave packet
propagates as a classical point-like particle, as long
as the wave packet is relativistic, $P_{0}/D \gg 1$.
Moreover, $\langle \mathbf{x} \rangle$ approaches $\mathbf{X}$ if $M_e$
is sufficiently large with respect to $D$.
In Sec.~\ref{sec:Quantum-scattering-Minkowski-spacetime},
we shall make numerical computations with $M_e/D = 10$.

\begin{table*}[t]
\vspace*{0mm}
\begin{center}
\caption{\label{tab:electron-wave-packets}%
Free electron wave packets: motion in three spatial dimensions.}\vspace*{4mm}
\renewcommand{\tabcolsep}{0.8pc} 
\renewcommand{\arraystretch}{1.8} 
\begin{tabular}{ |c|c|c|c| }
\hline
\hline
\multicolumn{4}{|c|}{Initial electron ($t=t_{i}$)} \\
\hline
&
$M_e/D=0.1$ &
$M_e/D=1$ &
$M_e/D=10$ \\
\hline
$\left|\langle x^{1} \rangle {-} X^{1} \right|\,D$ &
$\approx 2{\times}10^{-2}$ & $\approx 2{\times}10^{-3}$ & $\approx 2{\times}10^{-4}$ \\
$\left|\langle x^2 \rangle {-} X^2 \right|\,D$ &
$\approx 10^{-2}$ & $\approx 10^{-3}$ & $\approx 10^{-4}$ \\
$\left|\langle x^3 \rangle {-} X^3 \right|\,D$ & $0$ & $0$ & $0$ \\
\hline
$\left|\langle v^{1} \rangle {-} V^{1} \right|$ &
$\approx 4{\times}10^{-4}$ & $\approx 3{\times}10^{-5}$ & $\approx 10^{-5}$ \\
$\left|\langle v^2 \rangle {-} V^2 \right|$ &
$\approx 6{\times}10^{-4}$ & $\approx 4{\times}10^{-5}$ & $\approx 2{\times}10^{-5}$ \\
$\left|\langle v^3 \rangle {-} V^3 \right|$ & $0$ & $0$ & $0$ \\
\hline
\hline
\end{tabular}
\end{center}
\vspace*{0mm}
\begin{center}
\renewcommand{\tabcolsep}{0.8pc} 
\renewcommand{\arraystretch}{1.8} 
\begin{tabular}{ |c|c|c|c| }
\hline
\hline
\multicolumn{4}{|c|}{Final electron ($t=t_{f}$)} \\
\hline
&
$M_e/D=0.1$ &
$M_e/D=1$ &
$M_e/D=10$ \\
\hline
$\left|\langle x^{1} \rangle {-} X^{1} \right|\,D$ & $0$ & $0$ & $0$ \\
$\left|\langle x^2 \rangle {-} X^2 \right|\,D$ &
$\approx 2{\times}10^{-2}$ & $\approx 2{\times}10^{-3}$ & $\approx 2{\times}10^{-4}$ \\
$\left|\langle x^3 \rangle {-} X^3 \right|\,D$ & $0$ & $0$ & $0$ \\
\hline
$\left|\langle v^{1} \rangle {-} V^{1} \right|$ &
$\approx 8{\times}10^{-4}$ & $\approx 5{\times}10^{-5}$ & $\approx 2{\times}10^{-5}$ \\
$\left|\langle v^2 \rangle {-} V^2 \right|$ & $0$ & $0$ & $0$ \\
$\left|\langle v^3 \rangle {-} V^3 \right|$ & $0$ & $0$ & $0$ \\
\hline
\hline
\end{tabular}
\end{center}
\vspace*{-4mm}
\end{table*}

\subsection{Summary of initial and final conditions}
\label{subsec:Summary-initial-and-final-conditions}

\begin{figure}[t]   
\begin{center}
\includegraphics[scale=0.90]{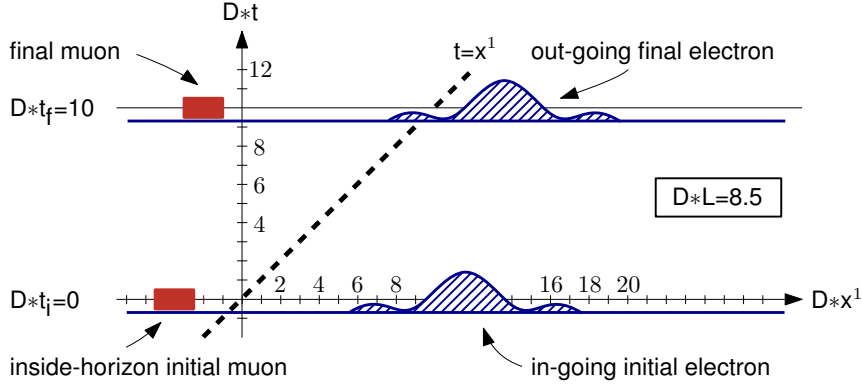}
\end{center}
\caption{Sketch of an elastic electron-muon scattering process
in Minkowski spacetime.
The solid-red (striped-blue) profiles show the spatial
support of the muon (electron) wave packets
at initial time $t=t_i=0$ and final time
$t=t_f=10/D$. The dashed line shows the projected
black-hole horizon (cf. the Penrose diagram of
Fig.~\ref{fig:bh-and-alcs}).
The initial muon is strictly localized \emph{within}
the projected black-hole horizon, while the initial
electron, localized with large probability outside
the projected black-hole horizon,
is strictly localized in momentum space with
finite support on \emph{negative} $Q^{1}$ momenta
(directed towards the black-hole center, left of the projected
black-hole horizon in this figure).
The final electron, localized with large probability outside
the projected black-hole horizon,
is strictly localized in momentum space with
finite support on \emph{positive} $Q^{1}$ momenta
(directed away from the black-hole center).
}\label{fig:sketch}
\end{figure}

The initial and final conditions for the muon wave packets
are given in \eqref{eq:ifc-muon}
and those for the electron wave packets
in \eqref{eq:ifc-electron}.
Both involve the momentum variance $D$,  whose inverse
sets the scale for all distances to be discussed later.
Figure~\ref{fig:sketch}
give a sketch of these initial and final conditions,
with crucial properties summarized in the caption.

As already remarked
in Secs.~\ref{subsec:Muon-wave-packets}
and \ref{subsec:Electron-wave-packets},
the initial and final free wave packets have been designed to
overlap at $t=5/D$
with center-of-mass coordinates $(0,\,0,\,0)$ for the muon
and $(L,\,0,\,0)$ for the electron.
These overlaps are needed to get a
significant value for the scattering probability amplitude
in Sec.~\ref{subsec:Numerical-results-probability-amplitude}.

\section{Quantum scattering in Minkowski spacetime}
\label{sec:Quantum-scattering-Minkowski-spacetime}

\subsection{Scattering probability amplitude without interactions}
\label{subsec:Probability-amplitude-without-interactions}

The probability amplitude corresponding to the scattering of
the muon and the electron in the absence of interactions
(coupling constant $\alpha = 0$) reads
\bsubeqs\label{eq:amplitude-no-interactions}
\beqa\label{eq:amplitude-no-interactions-a}
\mathcal{A}_{fi}^{(\alpha = 0)} &=&
\langle \psi_{X_{\mu,f},P_{\mu,f}}|\psi_{X_{\mu,i},P_{\mu,i}}\rangle
\;\langle \psi_{X_{e,f},P_{e,f}}|\psi_{X_{e,i},P_{e,i}}\rangle\,,
\eeqa
where
\beqa\label{eq:amplitude-no-interactions-b}
\hspace*{-5mm}
\langle \psi_{X_{\mu,f},P_{\mu,f}}|\psi_{X_{\mu,i},P_{\mu,i}}\rangle
&=& {\int}d^3\mathbf{x}\,
\big(\psi_{X_{\mu,f},P_{\mu,f}}(x)\big)^\dagger\;\psi_{X_{\mu,i},P_{\mu,i}}(x)
\nonumber\\[1mm]
\hspace*{-5mm}
&& \hspace*{-23.0mm}  =
{\int}\,\frac{d^3\mathbf{Q}}{(2\pi)^3}\,\big(F_{P_{\mu,f}}(\mathbf{Q})\big)_\mu
\big(F_{P_{\mu,i}}(\mathbf{Q})\big)_\mu\,e^{-iQ{\cdot}(X_{\mu,f}-X_{\mu,i})}\,,
\eeqa
\beqa
\label{eq:amplitude-no-interactions-c}
\hspace*{-5mm}
\langle \psi_{X_{e,f},P_{e,f}}|\psi_{X_{e,i},P_{e,i}}\rangle &=&
{\int}d^3\mathbf{x}\,
\big(\psi_{X_{e,f},P_{e,f}}(x)\big)^\dagger\;\psi_{X_{e,i},P_{e,i}}(x)
\nonumber\\[1mm]
\hspace*{-5mm}
&& \hspace*{-31.5mm}  =
{\int}\,\frac{d^3\mathbf{Q}}{(2\pi)^3}\,\big(F_{P_{e,f}}(\mathbf{Q})\big)_e
\big(F_{P_{e,i}}(\mathbf{Q})\big)_e\,e^{-iQ{\cdot}(X_{e,f}-X_{e,i})}
\;=\; 0\,.
\eeqa
\esubeqs
The last equality in \eqref{eq:amplitude-no-interactions-c}
is exact, because, for the chosen values of
$P_{e,i}^{1}$, $P_{e,f}^{1}$, and $D$,
\beqa
\text{rect}\left(\frac{Q^{1} - P_{e,f}^{1}}{2\,D}\right)\;
\text{rect}\left(\frac{Q^{1} - P_{e,i}^{1}}{2\,D}\right) &=&
0\,,
\eeqa
and similarly for the spatial $2$ direction.

Thus, we have from \eqref{eq:amplitude-no-interactions-a}
and \eqref{eq:amplitude-no-interactions-c},
in the absence of interactions,
\beqa
\mathcal{A}_{fi}^{(\alpha = 0)} &=& 0\,.
\eeqa
This result is, of course, as expected: the
initial electron has been designed to have only negative
$Q^{1}$ momenta, so that, without interactions, there is
zero overlap with a final electron having
only positive $Q^{1}$ momenta.

Any detection of a final electron with positive $Q^{1}$ momentum
requires nonzero momentum transfer, which, for the
case of the interacting quantum-electrodynamics theory,
traces back to the exchange of a virtual photon
(cf. the Feynman diagram in Fig.~\ref{fig:bh-and-alcs}).
Indeed, \emph{if} a final recoil electron is detected,
this means that there \emph{must}
have been an interaction
with a muon, at least for the setup considered.

We will now calculate the corresponding scattering probability
amplitude. Incidentally, we may call this process
``across-initial-muon-lightcone'' scattering,
as the recoil electron lies
across (outside) the outmost lightcone of the initial muon
wave packet; cf. Fig.~\ref{fig:sketch}.

\subsection{Scattering probability amplitude with interactions}
\label{subsec:Probability-amplitude-with-interactions}

The probability amplitude for the scattering
of the muon and the electron at the tree-level in
perturbation theory of quantum electrodynamics (QED)
reads~\cite{Feynman1949,Veltman1994}
\beqa\label{eq:tree-level-amplitude}
\mathcal{A}_{fi}^{\text{tree}} &=&
\frac{i\alpha}{4\pi^3}
{\int}d^4K\,\frac{\eta_{ab}\,J_e^{a}(K)J_\mu^b(K)}
{\left(K-P_{\mu,f}+P_{\mu,i}\right)^2+i\epsilon}\,,
\eeqa
where $\epsilon$ is a positive infinitesimal and
$\alpha \equiv e^2/4\pi$ the fine-structure constant.
The currents $J_e^{a}$ and $J_\mu^b$ in \eqref{eq:tree-level-amplitude}
are defined as follows:
\bsubeqs
\beqa
J_e^{a}(K) &\equiv& {\int}d^4x\,e^{-i(K-P_{\mu,f}+P_{\mu,i}){\cdot}x}\,
\nonumber \\[1mm] && \times\,
\Big[\big(\bar{\psi}_{X_{e,f},P_{e,f}}(x)\big)_e
\,\gamma^{a}\,
\big(\psi_{X_{e,i},P_{e,i}}(x)\big)_e\Big]\,,
\\[2mm]
J_\mu^b(K) &\equiv& {\int}d^4x\,e^{+i(K-P_{\mu,f}+P_{\mu,i}){\cdot}x}\,
\nonumber \\[1mm] && \times\,
\Big[\big(\bar{\psi}_{X_{\mu,f},P_{\mu,f}}(x)\big)_\mu
\,\gamma^b\,
\big(\psi_{X_{\mu,i},P_{\mu,i}}(x)\big)_\mu\Big]\,.
\eeqa
\esubeqs
The integrand of \eqref{eq:tree-level-amplitude}
is nonvanishing and  there is no obvious reason
for the integral to give zero generically,
that is, zero for \emph{all} values of
the distance parameter $L$
entering the boundary conditions \eqref{eq:ifc-electron}
of the initial and final electron wave packets,
for fixed boundary conditions \eqref{eq:ifc-muon} of the
initial and final muon wave packets.

The multiple integral \eqref{eq:tree-level-amplitude}
cannot be computed analytically.
Moreover, numerical calculation turns out to
be nontrivial and time-consuming.
We, therefore, make the approximation that
the momentum variance $D$
is much smaller than the particle masses, i.e., $M_e/D  \gg 1$
and $M_\mu/D  \gg 1$. We then have
\bsubeqs\label{eq:Je-and-Jmu-approx}
\beqa
\hspace*{-5mm}
J_e^{a}(K) &\approx&
E^{a}(K)
\nonumber \\[1mm]
\hspace*{-5mm}
&\equiv&
{\int}d^4x\,
\Big[\bar{e}_{X_{e,f},P_{e,f}}(x)
\,\gamma^{a}\,
e_{X_{e,i},P_{e,i}}(x)\Big]\,
e^{-i(K-P_{\mu,f}+P_{\mu,i}){\cdot}x}\,,
\\[2mm]
\hspace*{-5mm}
J_\mu^b(K) &\approx&
M^b(K)
\nonumber \\[1mm]
\hspace*{-5mm}
&\equiv&
{\int}d^4x\,
\Big[\bar{\mu}_{X_{\mu,f},P_{\mu,f}}(x)
\,\gamma^b\,
\mu_{X_{\mu,i},P_{\mu,i}}(x)\Big]\,
e^{+i(K-P_{\mu,f}+P_{\mu,i}){\cdot}x}\,,
\eeqa
\esubeqs
where $e_{X,P}(x)$ and $\mu_{X,P}(x)$ have been
defined in \eqref{eq:def-approx-muon-WF}
and \eqref{eq:def-approx-electron-WF}.
Explicitly, we obtain
\bsubeqs\label{eq:Ea-and-Mb-evaluated}
\beqa
\hspace*{-5mm}
E^{a}(K) &=& \frac{N_{e,f}N_{e,i}}{(\pi/D)^6}\,
\Big[\bar{u}(P_{e,f})\,\gamma^{a}\, u(P_{e,i})\Big]\,
e^{-i P_{e,f}{\cdot}X_{e,f}+i P_{e,i}{\cdot}X_{e,i}}
{\int}d^4x\,e^{-iK{\cdot}x}
\nonumber\\[1mm]
\hspace*{-5mm}
&& \hspace*{-18.0mm} \times
\prod_{i = 1}^{i = 3}
\text{sinc}\left[\frac{P_{e,f}^i\,\Delta{x}_{e,f}^0 -
P_{e,f}^0\,\Delta{x}_{e,f}^i}{P_{e,f}^0 \big/D}\right]\,
\text{sinc}\left[\frac{P_{e,i}^i\,\Delta{x}_{e,i}^0 -
P_{e,i}^0\,\Delta{x}_{e,i}^i}{P_{e,i}^0 \big/D}\right],
\eeqa
\beqa
\hspace*{-5mm}
M^b(K) &=& \frac{N_{\mu,f}N_{\mu,i}}{(2/D)^6}\,
\Big[\bar{u}(P_{\mu,f})\,\gamma^b\, u(P_{\mu,i})\Big]\,
e^{-iP_{\mu,f}{\cdot}X_{\mu,f}+iP_{\mu,i}{\cdot}X_{\mu,i}}
{\int}d^4x\,e^{+iK{\cdot}x}
\nonumber\\[1mm]
\hspace*{-5mm}
&& \hspace*{-18.0mm} \times
\prod_{i = 1}^{i =3}
\text{rect}\left[\frac{P_{\mu,f}^i\,\Delta{x}_{\mu,f}^0 -
P_{\mu,f}^0\,\Delta{x}_{\mu,f}^i}{2P_{\mu,f}^0\big/D}\right]\,
\text{rect}\left[\frac{P_{\mu,i}^i\,\Delta{x}_{\mu,i}^0 -
P_{\mu,i}^0\,\Delta{x}_{\mu,i}^i}{2P_{\mu,i}^0\big/D}\right].
\eeqa
\esubeqs
Both integrals over $x$ can be evaluated analytically,
the first with help of the formulae (3.742.6) and (3.742.8) in
Ref.~\cite{GradshteynRyzhik2007}.
The obtained expressions are, however, cumbersome
and will not be given here.

It should be noted that $E^{a}(K)$ has finite support. Specifically,
$E^{a}(K)$ is nonvanishing
if and only if
\bsubeqs
\beqa\hspace{-0mm}
K^0 &\in& \big({-}\mathcal{K},\,+\mathcal{K} \big)\,,
\\[2mm]
K^i/D &\in& \big({-}2,\,+2\big)\,,
\eeqa
where, in the first equation, we have taken into account the initial and
final conditions \eqref{eq:ifc-electron}
of the electron and where $\mathcal{K}$ is defined by
\beqa
\mathcal{K} &\equiv&
\left(\frac{P_{e,f}^{1}}{P_{e,f}^0} -
\frac{P_{e,i}^{1}}{P_{e,i}^0} +\frac{P_{e,i}^2}{P_{e,i}^0}
\right) D
\;\approx\; 2.393 \; D  \,.
\eeqa
\esubeqs

To summarize, we will evaluate the approximate amplitude
\beqa\label{eq:tree-level-amplitude-approx}
\mathcal{A}_{fi}^{\text{tree,\,approx}} &\equiv&
\frac{i\alpha}{4\pi^3}
{\int}d^4K\,\frac{\eta_{ab}\,E^{a}(K)\,M^b(K)}
{\left(K-P_{\mu,f}+P_{\mu,i}\right)^2+i\epsilon}\,,
\eeqa
where
$E^{a}(K)$ and $M^b(K)$ are given by \eqref{eq:Ea-and-Mb-evaluated}
with all integrals over $x$ performed analytically.
This approximation improves as the momentum variance $D$ drops to zero,
specifically
\beq
\label{eq:}
\mathcal{A}_{fi}^{\text{tree,\,approx}}\to \mathcal{A}_{fi}^{\text{tree}}
\,,\;\;\text{for}\;\;
D/M_e \to 0\,,
\eeq
where only the electron mass $M_{e}$ is shown for the $D$ limit,
as the muon mass $M_{\mu}$ is larger than $M_{e}$.
Note that $D$  is a naturally small parameter,
because it must be negligibly small with respect
to either the mass or the energy of the wave
packets for them to propagate approximately like classical particles
in special relativity.

\subsection{Numerical results for the scattering probability amplitude}
\label{subsec:Numerical-results-probability-amplitude}

For the numerical evaluation of the scattering probability
amplitude \eqref{eq:tree-level-amplitude-approx}, we take
the following parameter values:
\bsubeqs\label{eq:numerical-constants}
\beqa
M_e/D &=&  10
\,,
\\[2mm]
M_\mu/D &=& 2070
\,,
\\[2mm]
\alpha &=& 1/137\,,
\eeqa
\esubeqs
where the electron mass $M_{e}$
and the approximate muon
mass $M_{\mu} \equiv 207\,M_{e}$
are considered to be fixed and $D$ is the momentum variance
of the assumed wave packets of the initial and final particles.
As a start, we have compared,
for the numerical values \eqref{eq:numerical-constants} and
the parameter value $L=8.5/D$
[entering the boundary conditions
of the scattering process considered],
the numerical result for the
$J_e(K)\cdot J_\mu(K)$ numerator in the integrand
of \eqref{eq:tree-level-amplitude} with
the analytic result for the $E(K)\cdot M(K)$ numerator
in the integrand of \eqref{eq:tree-level-amplitude-approx},
and find excellent agreement.

\begin{figure}[t]  
\begin{center}
\includegraphics[scale=1.1]{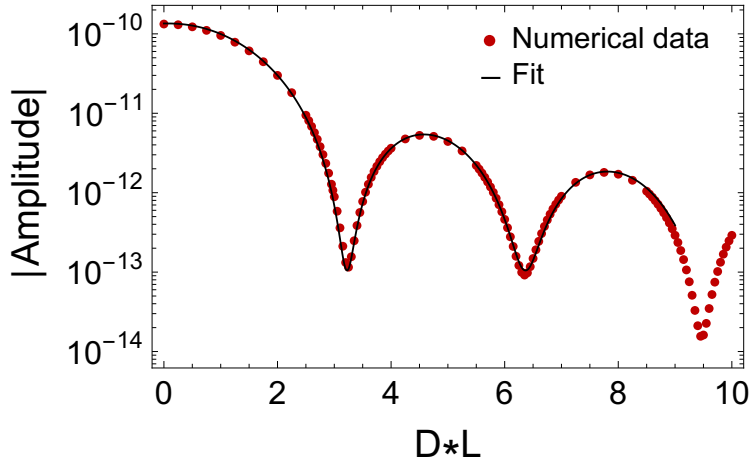}
\end{center}
\vspace*{-4mm}
\caption{Absolute value of the scattering probability amplitude
$\mathcal{A}_{fi}^\text{tree,\,approx}$
from \eqref{eq:tree-level-amplitude-approx}
as a function of the distance $L$ entering
the boundary conditions \eqref{eq:ifc-muon} and
\eqref{eq:ifc-electron} [the initial and final wave functions 
involve the momentum variance $D$]. Numerical parameters
\eqref{eq:numerical-constants} are used.
The fit for $D\,L \in [0,\,9]$
is given by expression \eqref{eq:A-tree-approx-fit-even}.}
\label{fig:amplitude}
\end{figure}

The results for the numerical evaluation
of \eqref{eq:tree-level-amplitude-approx}
are shown in Fig.~\ref{fig:amplitude}.
The rough order of magnitude of $10^{-11}$
for the amplitude in Fig.~\ref{fig:amplitude}
can be understood as follows:
the prefactor on the right-hand side
of \eqref{eq:tree-level-amplitude-approx}
gives a factor of order $10^{-4}$,
the numerator at $K^\mu=0$ a factor of order $1$,
and the denominator at $K^\mu=0$ a factor of order $10^{-7}$.
Comparing different numerical calculations, we estimate the
numerical accuracy of the results shown in Fig.~\ref{fig:amplitude}
to be approximately $10^{-13}$.

\vspace*{0mm}

Purely empirically, we can fit the Fig.~\ref{fig:amplitude}
numerical results for $D\,L \in [0,\,9]$
by the following even function:%
\bsubeqs\label{eq:A-tree-approx-fit-even}
\beqa\label{eq:A-tree-approx-fit-function-even}
\hspace*{-4mm}
\Big|\mathcal{A}_{fi}^\text{tree,\,approx}\big(D\,L\big)\Big|^\text{(fit)}
&=&
a_{0} \,\Big|\text{sinc}\left(D\,L- D\,\widetilde{\ell}_{0}\right)\Big|^{p_{0}}
+b\big(D\,L\big) \,,
\\[2mm]
\widetilde{\ell}_{0} &=&
\frac{L}{\sqrt{\ell_{0}^2+L^2}}\;\ell_{0}\,,
\\[2mm]
\label{eq:A-tree-approx-fit-b0-even}
b\big(D\,L\big)  &=& \frac{b_{0}}{1+b_{2}\,\big(D\,L\big)^{2}}\,,
\\[2mm]\hspace*{-0mm}
\label{eq:A-tree-approx-fit-1st-parameters-even}
\big\{a_{0},\, \ell_{0},\, p_{0} \big\}
&\approx&
\left\{
1.35 {\times} 10^{-10},\, 0.084/D,\,2.12 \right\},
\\[2mm]\hspace*{-0mm}
\label{eq:A-tree-approx-fit-2nd-parameters-even}
\big\{b_{0},\,  b_{2}\big\}
&\approx&
\left\{1.05 {\times} 10^{-13},\, 0 \right\},
\eeqa
\esubeqs
where the function `$\text{sinc}$' is defined by \eqref{eq:def-sinc}.
A positive value of $b_{2}$ can perhaps be determined from better
numerical data of the third dip at
\mbox{$D\,L \sim 3\,\pi$.}

\newpage 
The nonzero result for the scattering probability amplitude
(Fig.~\ref{fig:amplitude})
implies that Minkowski-spacetime across-initial-muon-lightcone scattering
as shown in Fig.~\ref{fig:sketch} can take place in QED.
As mentioned in Sec.~\ref{sec:Introduction},
this implies, according to the Einstein Equivalence Principle,
that across-horizon scattering is operative in a black-hole spacetime
and we will discuss this further in the next section.

\section{Across-horizon scattering for a massive black hole}
\label{sec:Across-horizon-scattering-for-massive-black-hole}

In the present article, we have established,
by explicit calculation, the nonzero probability for detecting a recoil
electron in a special setup of elastic electron-muon scattering
in Minkowski spacetime, where
the recoil electron is detected outside the light-cone of
the initial muon (Fig.~\ref{fig:sketch}).
The setup is such that there is zero probability for detecting
a recoil electron if there is no interaction taking place,
which is the case if, for example, the initial muon is stopped
by a closed shutter.

The Minkowski-spacetime result of a nonzero probability
amplitude for a recoil electron (Fig.~\ref{fig:amplitude})
provides the cap-stone of our previous
argument~\cite{EmelyanovKlinkhamer2018}
to establish the possibility of information transfer
out of a Schwarzschild black hole.
Indeed, a near-horizon spacetime region is approximately flat for a
very massive black hole and this spacetime region
with local inertial coordinates
can be described by a patch of Minkowski spacetime
(see Fig.~2 in Ref.~\cite{EmelyanovKlinkhamer2018}),
where the original black-hole horizon projects on part of the
light-cone in Minkowski spacetime (dashed line in Fig.~\ref{fig:sketch}).
By the Einstein Equivalence Principle,
the physics in this Minkowski-spacetime patch
is described by standard QED~\cite{Feynman1949,Veltman1994},
if we consider electrically charged elementary particles and photons
and omit
the weak and strong interactions.
Specifically, the flat-spacetime calculation of
Sec.~\ref{sec:Quantum-scattering-Minkowski-spacetime}
is relevant for elastic electron-muon scattering
in a near-horizon spacetime region
of a Schwarzschild black hole with a sufficiently large mass
($M_\text{BH} \gg M_{P}^2\,c/D$,
where $M_{P}^2 \equiv \hbar c/G$
is the square of the Planck mass and
$\hbar/D$ the characteristic quantum-particle size).
See also App.~B of Ref.~\cite{EmelyanovKlinkhamer2018}
for further relevant numbers.

In Ref.~\cite{EmelyanovKlinkhamer2018}, we have proposed a
\textit{Gedankenexperiment}
that relies on the nonzero probability for a recoil electron,
which is precisely what we have calculated in the present paper.
A brief summary of this \textit{Gedankenexperiment} is as follows.
Two experimentalists, Castor and Pollux, meet outside the
black-hole horizon to fix the procedure
and also to establish a list of questions for Castor to answer.
Castor then moves inside
the black-hole horizon and starts the experiment at an agreed moment.
While Pollux on the outside definitely sends out an appropriate
bunch of electrons, Castor on the inside decides to
send an appropriate bunch of muons
if his answer to the first question is affirmative
or decides not to send an appropriate bunch of muons
if his answer to the first question is negative.
Castor  writes ``yes'' in his message-book if he sends
the muons and ``no'' if he does not.
The detector of Pollux is designed to record recoil electrons and
Pollux writes ``YES'' in his logbook if there is at least one
detected recoil electron and ``NO'' if there are no recoil
electrons whatsoever. Hence, Castor's yes/no
answer to the first question is transmitted to Pollux,
who reads YES/NO in his logbook. Castor and Pollux
deal with further questions in the same way.
Additional details and refinements can be found
in Sec. 3 of Ref.~\cite{EmelyanovKlinkhamer2018}.

To summarize, we have established that it is, in principle, possible to
transmit a message (in binary code) from inside the black-hole horizon
to outside the black-hole horizon by use of a quantum scattering process
(Fig.~\ref{fig:bh-and-alcs}).
Another quantum process is, of course,
spontaneous pair production,
which plays a crucial role for the Hawking radiation
of a black hole~\cite{Hawking1975}.
But it appears that this spontaneous pair-production
process, in its simplest form, cannot be used to
send a message from inside the black-hole horizon to outside.
Still, even though the quantum scattering process
can, in principle,
be used to send such a message, it is not clear
what this result implies for the so-called information-loss
problem of black-hole physics~\cite{UnruhWald2017}.

\section*{\hspace*{-4.5mm}Acknowledgments}

FRK gratefully remembers Tini Veltman for the many
discussions we had on quantum field theory and gravity.


\end{document}